\documentclass[lettersize,journal]{IEEEtran}
\usepackage{cite}
\usepackage{amsmath,amssymb,amsfonts}
\usepackage{algorithmic}
\usepackage{graphicx}
\usepackage{textcomp}
\usepackage{xcolor}
\usepackage{subcaption}
\usepackage{caption}
\DeclareCaptionJustification{justified}{\justifying}
\usepackage[raggedrightboxes]{ragged2e}
\usepackage{algorithm}
\usepackage{algorithmic}
\usepackage{float}
\usepackage{multirow}
\usepackage{booktabs} % for better table lines
\usepackage{tabularx}
\usepackage{multirow}
\usepackage[colorlinks=true, linkcolor=blue, citecolor=blue, urlcolor=blue]{hyperref}
\begin{document}

\title{Neural B-Frame Coding: Tackling Domain Shift Issues with Lightweight Online Motion Resolution Adaptation}

\author{Sang NguyenQuang,~\IEEEmembership{Student Member,~IEEE}, Xiem HoangVan, and~Wen-Hsiao Peng,~\IEEEmembership{Fellow,~IEEE}
        % <-this % stops a space
% \thanks{This paper was produced by ...}% <-this % stops a space
% \thanks{Manuscript received ..; revised ...}
}

\markboth{}%
{Shell \MakeLowercase{\textit{et al.}}: A Sample Article Using IEEEtran.cls for IEEE Journals}


\maketitle

% \input{chapters/0_Abstract}
\begin{abstract}
Learned B-frame codecs with hierarchical temporal prediction often encounter the domain-shift issue due to mismatches between the Group-of-Pictures (GOP) sizes for training and testing, leading to inaccurate motion estimates, particularly for large motion. A common solution is to turn large motion into small motion by downsampling video frames during motion estimation. However, determining the optimal downsampling factor typically requires costly rate-distortion optimization. This work introduces lightweight classifiers to predict downsampling factors. These classifiers leverage simple state signals from current and reference frames to balance rate-distortion performance with computational cost. Three variants are proposed: (1) a binary classifier (Bi-Class) trained with Focal Loss to choose between high and low resolutions, (2) a multi-class classifier (Mu-Class) trained with novel soft labels based on rate-distortion costs, and (3) a co-class approach (Co-Class) that combines the predictive capability of the multi-class classifier with the selective search of the binary classifier. All classifier methods can work seamlessly with existing B-frame codecs without requiring codec retraining. Experimental results show that they achieve coding performance comparable to exhaustive search methods while significantly reducing computational complexity. The code is available at: \href{https://github.com/NYCU-MAPL/Fast-OMRA.git}{https://github.com/NYCU-MAPL/Fast-OMRA.git}.
\end{abstract}

\section{Introduction}
\label{sec:Introduction}
\IEEEPARstart{L}{earned} video coding has achieved promising results, with some methods~\cite{dcvcdc, dcvcfm, dcvcsdd, dcvcrt} surpassing H.266/VVC~\cite{overview_vvc}. However, most research focuses on P-frames, while B-frame coding~\cite{TLZMC, BEPIC, bcanf, maskcrttcsvt} remains underexplored.

\begin{figure}[t]
    \centering
    \vspace{-1em}
    \centerline{\includegraphics[width=0.37\textwidth]{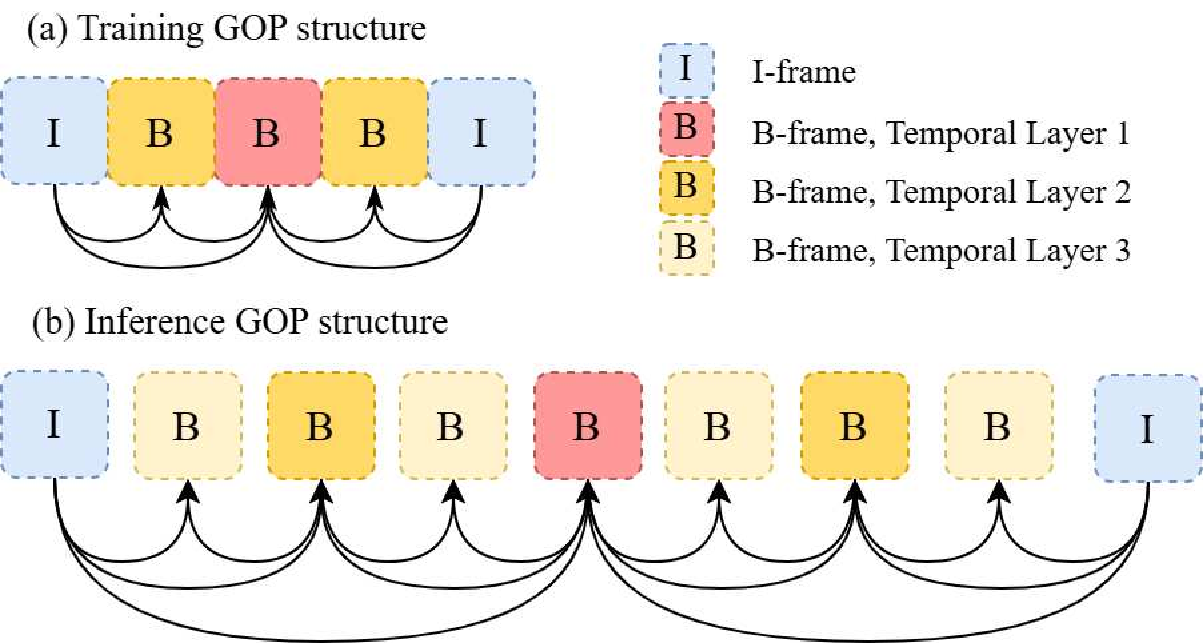}}
    \caption{The GOP structure in the training process (a) consists of only five frames, whereas the inference process (b) can extend up to 32 frames.}
    \vspace{-1.0em}
    \label{fig:domain-shift}
\end{figure}

Learned B-frame coding is specifically designed to exploit both past and future temporal information for enhanced coding efficiency. But most existing approaches fail to outperform learned P-frame coding. One primary reason for this performance gap is the domain-shift issue, which is first identified in~\cite{bcanf}. As illustrated in Fig.~\ref{fig:domain-shift}, it arises when B-frame codecs are trained on short video sequences with small Groups of Pictures (GOPs), but are subsequently applied to encode large GOPs during inference. The mismatch between training and testing GOP sizes leads to suboptimal coding performance.

To address the domain-shift issue, a straightforward solution is to train the codec on long video sequences~\cite{longvideocoding}. However, this method is time-consuming. As an alternative solution, Gao~\emph{et al.}~\cite{OMRA} propose an Online Motion Resolution Adaptation (OMRA) method for neural B-frame coding that avoids re-training the codec. This approach adaptively downsamples both coding and reference frames for motion estimation, effectively converting large motion into small motion when necessary. Despite its simplicity, OMRA achieves substantial improvements in coding performance. As a concurrent work, Yilmaz~\emph{et al.}~\cite{Murat_ICIP} propose using warped frame quality to predict the optimal downsampling factor, leveraging flow fields at different resolutions to guide deformable convolution for improved motion estimation. Bilican~\emph{et al.}~\cite{Murat_OJSP} further extend the idea to P-frame coding.

Based on OMRA and the MaskCRT B-frame codec~\cite{maskcrttcsvt} (hereafter referred to as MaskCRT-B), our previous work~\cite{FastOMRA} introduces a fast decision strategy, where a lightweight neural network is utilized to predict the downsampling factor. This approach achieves an 11\% bitrate reduction while maintaining computational complexity lower than that of MaskCRT-B. In an effort to extend~\cite{FastOMRA}, this work presents a comprehensive discussion of the domain-shift problem in Neural B-frame coding and introduces a new variant Co-Class as an additional lightweight neural prediction approach for low-complexity motion resolution adaptation. It integrates multi-class classifier predictions with a selective search step to enhance prediction accuracy. Table~\ref{tab:method_compare} summarizes the key differences between this extended work and related prior research.

Specifically, this work has the following contributions:
\begin{itemize}
    \item First, we present Fast-OMRA, a lightweight online motion resolution adaptation solution that employs neural prediction for resolution determination.
    \item Second, we introduce the Co-Class variant, which completes the Fast-OMRA framework along with Bi-Class and Mu-Class presented in~\cite{FastOMRA}.
    \item Third, we conduct extensive experiments to validate the proposed Fast-OMRA architectures.
\end{itemize}
\begin{table}[t!]
\caption{Our Fast-OMRA versus prior research.}
\scriptsize
\centering
\begin{tabular}{ccccc}
\hline
Method                                                      & \begin{tabular}[c]{@{}c@{}}Coding\\ Scheme\end{tabular} & \begin{tabular}[c]{@{}c@{}}Codec\\ Specific\end{tabular} & \begin{tabular}[c]{@{}c@{}}Search\\ Method\end{tabular} & \begin{tabular}[c]{@{}c@{}}Search\\ Criteria\end{tabular}                                                       \\ \hline
ICIP'24~\cite{Murat_ICIP}                                                         & B-frame                                                 & Yes                                                      & Exhaustive                                               & Warped frame quality \\ \hline
OJSP'25~\cite{Murat_OJSP}                                                         & P-frame                                                 & No                                                       & Exhaustive                                               & Warped frame quality \\ \hline
OMRA~\cite{OMRA}                                                         & B-frame                                                 & No                                                       & Exhaustive                                               & RD-cost                                                        \\ \hline
\begin{tabular}[c]{@{}c@{}}Bi-Class\\ Fast-OMRA~\cite{FastOMRA}\end{tabular} & B-frame                                                 & No                                                       & Selective                                                & Warped frame quality \\
\begin{tabular}[c]{@{}c@{}}Mu-Class\\ Fast-OMRA~\cite{FastOMRA}\end{tabular} & B-frame                                                 & No                                                       & X                                                        & \begin{tabular}[c]{@{}c@{}}X\end{tabular} \\

\begin{tabular}[c]{@{}c@{}}Co-Class\\ Fast-OMRA\end{tabular} & B-frame                                                 & No                                                       & Selective                                                & Warped frame quality \\ \hline
\end{tabular}
\vspace{-1.5em}
\label{tab:method_compare}
\end{table}
\vspace{-1em}

\begin{figure*}[t]
    \centering
    \centerline{\includegraphics[width=0.95\textwidth]{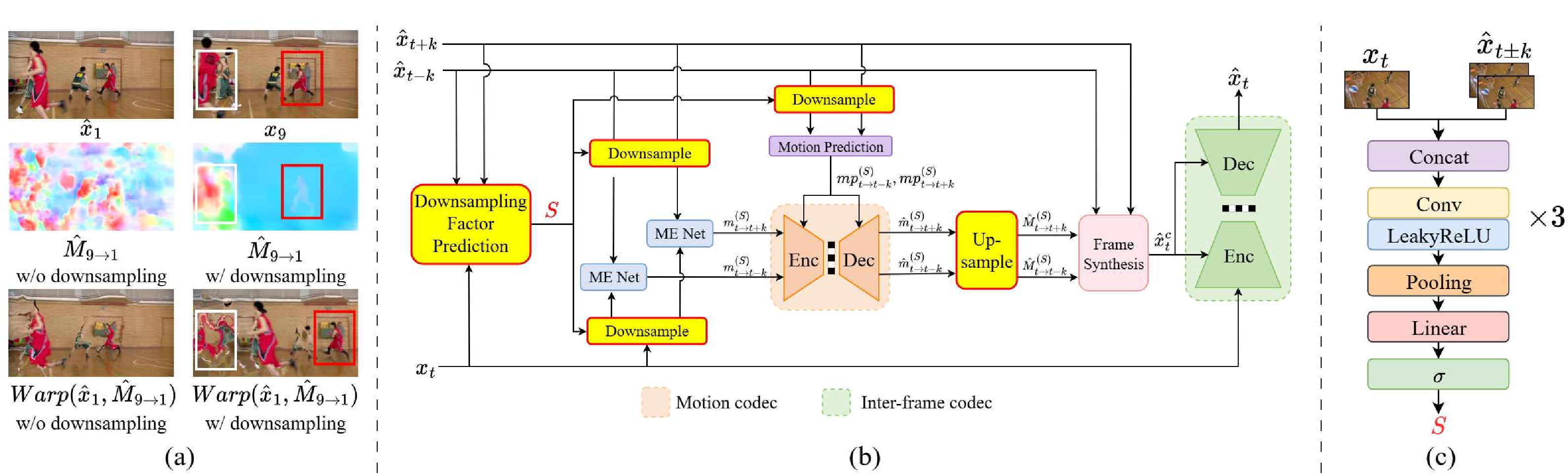}}
    \caption{(a) Motion estimation and temporal warping with and without downsampling, (b) Online Motion Resolution Adaptation for neural B-frame coding, and (c) classification network architecture.}
    \label{fig:omra}
    \vspace{-1.5em}
\end{figure*}
\section{The Domain-Shift Issue and OMRA}
As presented in~\cite{OMRA}, learned B-frame codecs with hierarchical temporal prediction are often trained on short videos but have to encode videos with large GOP sizes in practice. Consequently, the encoder struggles with motion estimation for video frames in low temporal layers, where coding and reference frames are temporally distant. This challenge is especially pronounced in videos with fast, complex motion. As depicted in Fig.~\ref{fig:omra}(a), it can be observed that the warped frame generated without downsampling fails to capture meaningful information. In contrast, when downsampling is applied, certain regions within the warped frame retain informative content that can be leveraged for the conditioning signal, as highlighted in the red boxes. However, this method still cannot handle areas with highly complex motion, such as regions where objects appear in the current frame but are absent from the reference frames, as shown in the white boxes.

Fig.~\ref{fig:omra}(b) illustrates a neural B-frame codec incorporating Online Motion Resolution Adaptation (OMRA). The encoder first determines the optimal downsampling factor $S$ from the set $\left\{ 1,2,4,8\right\}$. The current frame $x_t$ and reference frames $\hat{x}_{t-k},\hat{x}_{t+k} \in \mathbb{R}^{3 \times W \times H}$ are then downsampled accordingly to $x^{(S)}_t, \hat{x}^{(S)}_{t-k}, \hat{x}^{(S)}_{t+k} \in \mathbb{R}^{3 \times \frac{W}{S} \times \frac{H}{S}}$. Motion estimation and motion prediction are performed on these downsampled frames. The resulting optical flow maps $m^{(S)}_{t\to{t}-k}, m^{(S)}_{t\to{t}+k} \in \mathbb{R}^{2 \times \frac{W}{S} \times \frac{H}{S}}$ are then compressed by leveraging the predicted motion $mp^{(S)}_{t\to{t}-k}, mp^{(S)}_{t\to{t}+k} \in \mathbb{R}^{2 \times \frac{W}{S} \times \frac{H}{S}}$ as the condition signal. At the decoder, the reconstructed optical flow maps $\hat{m}^{(S)}_{t\to{t}-k}, \hat{m}^{(S)}_{t\to{t}+k} \in \mathbb{R}^{2 \times \frac{W}{S} \times \frac{H}{S}}$ are upsampled to full resolution as $\hat{M}_{t\to{t}-k}, \hat{M}_{t\to{t}+k} \in \mathbb{R}^{2 \times W \times H}$, and fed into the Frame Synthesis module for motion compensation.

% \subsection{Determining downsampling factor}
Ideally, the downsampling factor $S$ of each B-frame should be selected from the pre-defined set $\left\{ 1,2,4,8\right\}$ by minimizing the rate-distortion cost:
\begin{equation}
\label{equ:rd_loss_P}
 %\underset{S\in \left\{1,2,4,8\right\}}{\min} 
  \lambda \cdot D(x_t, \hat{x}_t) + r_t,
\end{equation}
where $D, r_t, \lambda$ are the distortion between the coding $x_t$ and reconstructed $\hat{x}_t$ frames, the total bitrate needed to encode $x_t$, and the Lagrange multiplier, respectively.

\begin{table}[t]
\caption{BD-rate savings (\%) and encoding complexity in kilo multiply-accumulate operations per pixel (kMAC/pixel) of OMRA and its variants.}
\scriptsize
\begin{tabular}{cccccc}
\hline
\multirow{2}{*}{Method} & \multicolumn{4}{c}{BD-rate}        & \multirow{2}{*}{kMAC/pixel} \\ \cline{2-5}
                        & HEVC-B & UVG   & MCL-JCV & Average &                             \\ \hline
MaskCRT-B         & 0.0    & 0.0   & 0.0     & 0.0     & 2033                        \\
OMRA                    & -9.7  & -15.0 & -16.0   & -13.6   & 5863                        \\
OMRA-MEMC               & -8.3   & -13.5 & -11.7   & -11.2   & 2235                        \\
OMRA-MEMC*              & -8.3   & -13.4 & -9.7    & -10.5   & 2131                        \\ \hline
\end{tabular}
\label{tab:RDC_compare}
\vspace{-1.7em}
\end{table}
However, performing full encoding and decoding four times in the present case (and even more when there are more downsampling factors to evaluate) is computationally expensive, making this approach impractical despite its optimal performance. Therefore, a fast search algorithm is needed to reduce complexity and accelerate the selection of the downsampling factor. OMRA~\cite{OMRA} shows that applying temporal warping with an optical flow map derived from the optimal downsampling factor results in higher-quality warped frames. To reduce complexity, OMRA-MEMC~\cite{FastOMRA} predicts the downsampling factor by assessing the prediction error between the original and warped frames without requiring encoding and decoding. OMRA-MEMC*~\cite{FastOMRA} further optimizes this process by eliminating the evaluation of multiple downsampling factors for video frames in the highest temporal layer, where the temporal prediction distance is minimal. 

Table~\ref{tab:RDC_compare} presents a comparison of the BD-rate savings~\cite{bdrate} and encoding kMAC/pixel among the competing methods, using MaskCRT-B as the anchor. The per-dataset BD-rate is evaluated by averaging the BD-rates of individual videos in a dataset. It is seen that integrating OMRA with full rate-distortion optimization (RDO) significantly increases the computational cost. In contrast, OMRA-MEMC and OMRA-MEMC* introduce modest complexity overhead, as they perform motion estimation without requiring motion and inter-frame coding. Regarding BD-rate savings, OMRA-MEMC and OMRA-MEMC* exhibit slight degradation compared to OMRA, with average BD-rate reductions of 2\% and 3\%, respectively.

\vspace{-1.0em}

\section{Proposed Method: Fast-OMRA}
In order to reduce both the number of search steps and encoding complexity of OMRA, we propose Fast-OMRA, a novel method that employs a lightweight classification network to predict the downsampling factor for each B-frame and its reference frames. Two variants of this approach incorporate additional search steps to refine the selection of the downsampling factor $S$. As illustrated in Fig.~\ref{fig:omra}(c), the network architecture of Fast-OMRA concatenates the current and reference frames, processing them through three convolutional layers, followed by a pooling layer and a linear layer to generate the final prediction.
\vspace{-1.0em}

\subsection{Bi-Class and Mu-Class Fast-OMRA}
\label{ssec:analysis_fastOMRA}
In~\cite{FastOMRA}, Fast-OMRA is introduced with two variants: Bi-Class and Mu-Class. Bi-Class Fast-OMRA uses a neural network as a binary classifier to determine whether the coding frame $x_t$ and its reference frames $\hat{x}_{t-k},\hat{x}_{t+k}$ should undergo motion estimation and motion coding in full resolution or low resolution. If the model predicts full resolution, $S$ is set to 1. Otherwise, a warped frame quality-based search process, similar to OMRA-MEMC, is performed over the candidate set $\left\{2, 4, 8\right\}$. In contrast, Mu-Class Fast-OMRA minimizes search complexity by a multi-class classifier that directly outputs $S$ without requiring any search process. 

Specifically, Bi-Class is trained using Focal Loss, with each temporal layer having its own classification network. The training objective of Bi-Class is given by:
\begin{equation}
\label{equ:focal_loss}
L_{Bi} = \alpha_{t} \cdot \left ( 1-p_{t} \right )^\gamma \cdot CE(p_t, L_{hard}),
\end{equation}
where $\alpha_t$ is used to balance class distributions, $(1-p_t)^\gamma$ discounts well-classified samples, $\gamma$ is set empirically to 2, $L_{hard}$  is the one-hot ground truth, and CE represents the cross-entropy loss. $p_t$ is equal to the classifier's output $p$ if the label is 1, otherwise $1-p$.

Mu-Class uses soft labels with entropy weighting to handle ambiguous samples. The soft label is:
\begin{equation}
\label{equ:focal_loss}
L_{soft} = \frac{e^{\lambda a_{i}}}{\sum e^{\lambda a_{i}}},
\end{equation}
\begin{equation}
\label{equ:focal_loss}
 a_{i}= \frac{RD_{max}-RD_{i}}{RD_{max}},
\end{equation}
where $RD_i$ is the Rate-Distortion (RD) cost for downsampling factor $i$, $RD_{max}$ is the largest RD cost among the four downsampling factors, and $\lambda$ is set to 10. Recognizing that ambiguous samples with similar RD costs across downsampling factors can confuse the model, an entropy measure of $L_{soft}$ is used to distinguish ambiguous samples from distinct ones. In the ambiguous case, $L_{soft}$ approaches 0.25 for each class ($Entropy \approx 2$), and these samples are discounted during training. We construct the training objective as follows:
\begin{equation}
\label{equ:focal_loss}
L_{Mu} = \left ( 2-Entropy(L_{soft}) \right ) \cdot CE(p_t, L_{soft}),
\end{equation}
where $L_{soft}$ is the softened ground truth label.

\vspace{-1.0em}
% \subsection{Proposed Fast-OMRA++}
% \label{ssec:proposed}
\subsection{Co-Class Fast-OMRA}
\begin{figure}[t]
    \centering
    % \vspace{-1em}
    \centerline{\includegraphics[width=0.45\textwidth]{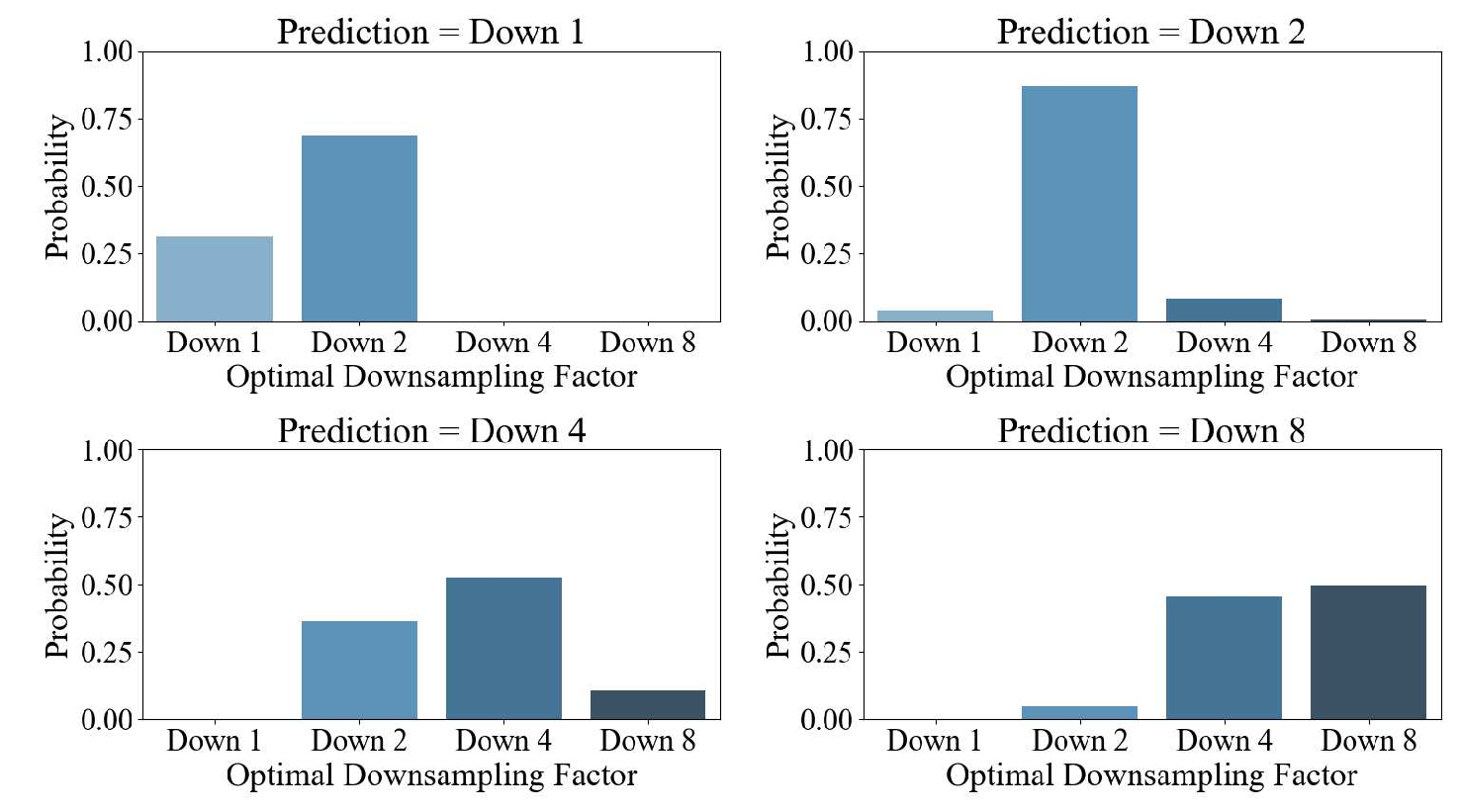}}
    \caption{The conditional probability distributions of ground-truth downsampling factors given the predicted outcomes of Mu-Class on the BVI-DVC and TVD datasets.}
    \label{fig:confusion_BVI}
    \vspace{-1.5em}
\end{figure}
\begin{table}[t]
\centering
\caption{Candidate sets of Co-Class Fast-OMRA.}
\scriptsize
\begin{tabular}{ccccc}
\hline
Mu-Class Prediction & 1        & 2        & 4        & 8        \\
Candidate set $S_c$ & \{1, 2\} & \{2, 4\} & \{2, 4\} & \{4, 8\} \\ \hline
\end{tabular}
\label{tab:candidate_co_class}
\vspace{-1.7em}
\end{table}
Additionally, this work introduces Co-Class Fast-OMRA, an algorithm that combines the predictive capabilities of Mu-Class with the selective search process of Bi-Class to enhance the prediction accuracy while maintaining low encoding complexity. To achieve this, Mu-Class is first used to predict a specific downsampling factor $S$. Then, its neighboring downsampling factors in a candidate set $S_c$ are evaluated to improve prediction accuracy. Specifically, $S_c$ is established by analyzing the conditional probability distributions of ground-truth downsampling factors given the prediction outcomes. For example, in Fig.~\ref{fig:confusion_BVI}, when the predicted outcome is Down 1, both Down 2 (68.75\%) and Down 1 (31.25\%) are highly probable ground-truth downsampling factors, and should be evaluated. Based on this analysis on video sequences in the BVI-DVC~\cite{BVI} and TVD~\cite{TVD} datasets, the candidate set is determined and summarized in Table~\ref{tab:candidate_co_class} for each possible predicted outcome.

For each candidate $S$ in the candidate set $S_{c}$, the coding frame and reference frames are downsampled by $S$ as $x^{(S)}_t, \hat{x}^{(S)}_{t-k}, \hat{x}^{(S)}_{t+k} \in \mathbb{R}^{3 \times \frac{W}{S} \times \frac{H}{S}}$ before performing motion estimation. The resulting optical flow maps $m^{(S)}_{t\to{t}-k}, m^{(S)}_{t\to{t}+k} \in \mathbb{R}^{2 \times \frac{W}{S} \times \frac{H}{S}}$ are then super-resolved to the original spatial resolution. Afterwards, temporal warping is applied to obtain the warped frames $x_{t \to t-k}^{(S)},x_{t \to t+k}^{(S)} \in \mathbb{R}^{3 \times W \times H}$. The optimal downsampling factor $S$ is the one that minimizes temporal prediction errors.

\vspace{-1.0em}

\begin{figure*}[t]
    \begin{center}
    \vspace{-1.3em}
    \begin{subfigure}{0.3\linewidth}
        \centering
        \includegraphics[width=\linewidth]{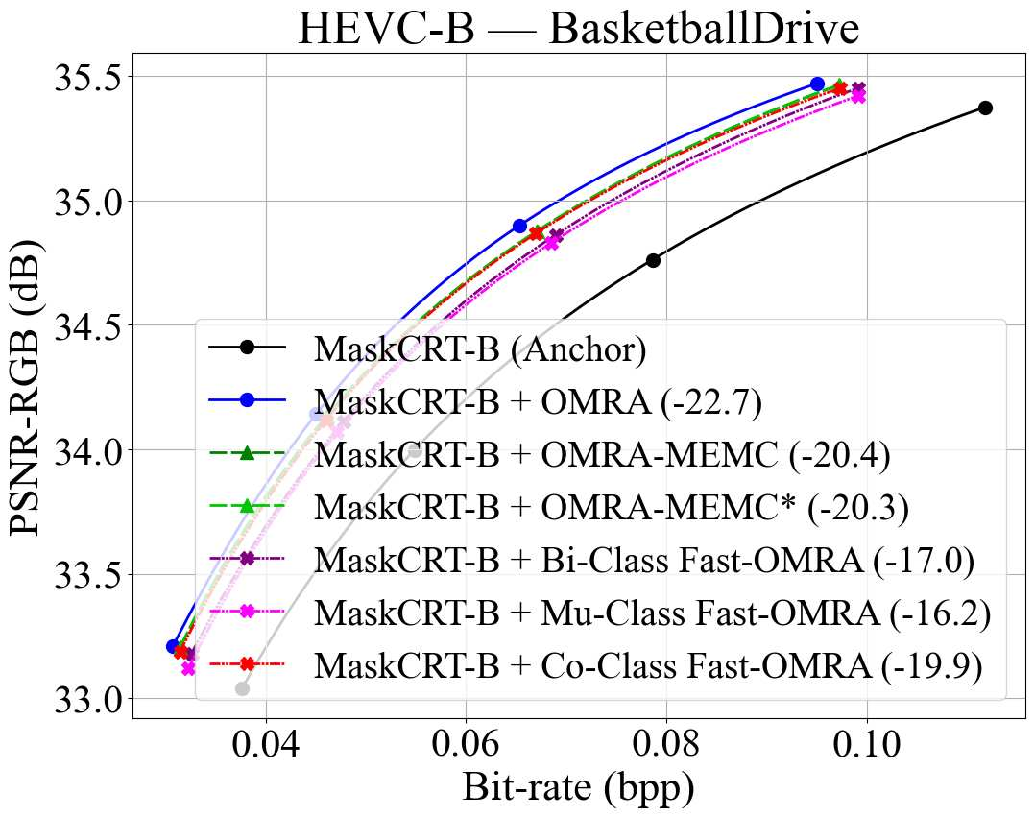}
        % \caption{\centering HEVC Class B}
        \label{fig:hevcb}
    \end{subfigure}
    \begin{subfigure}{0.3\linewidth}
        \centering
        \includegraphics[width=\linewidth]{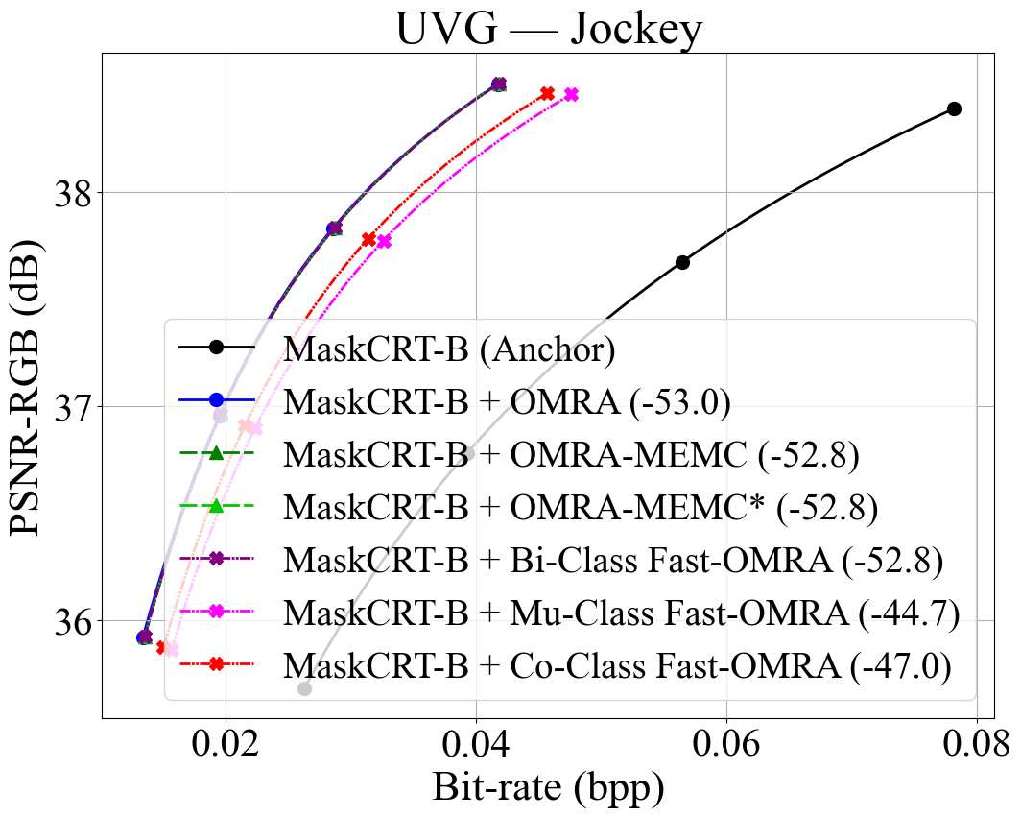}
        % \caption{\centering UVG}
        \label{fig:uvg}
    \end{subfigure}
    \begin{subfigure}{0.3\linewidth}
        \centering
        \includegraphics[width=\linewidth]{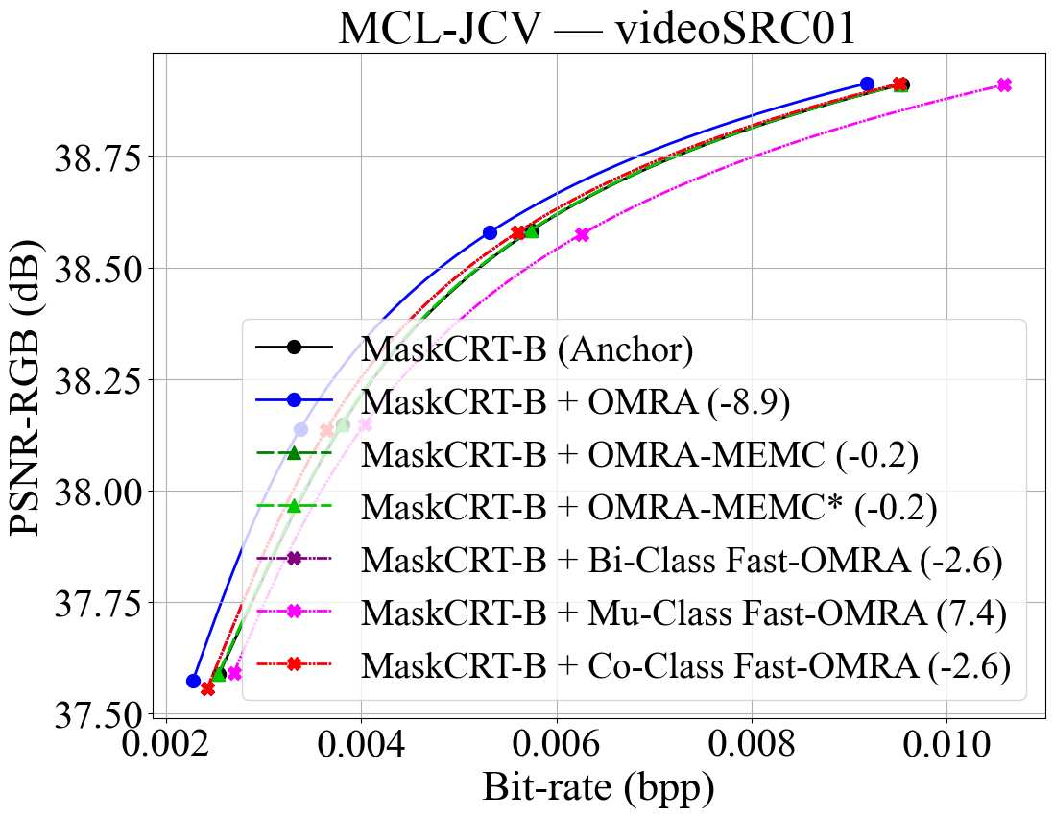}
        % \caption{\centering MCL-JCV}
        \label{fig:mcl}
    \end{subfigure}
    \vspace{-1.3em}
    \caption{The rate-distortion performance comparison for BasketballDrive, Jockey and videoSRC01 sequences.}
    \label{fig:RD_seq}
    \end{center}
    \vspace{-1.0em}
\end{figure*}
\begin{figure*}[t]
    \begin{center}
    \vspace{-1.1em}
        \begin{subfigure}{0.64\linewidth}
        \centering
        \includegraphics[width=\linewidth]{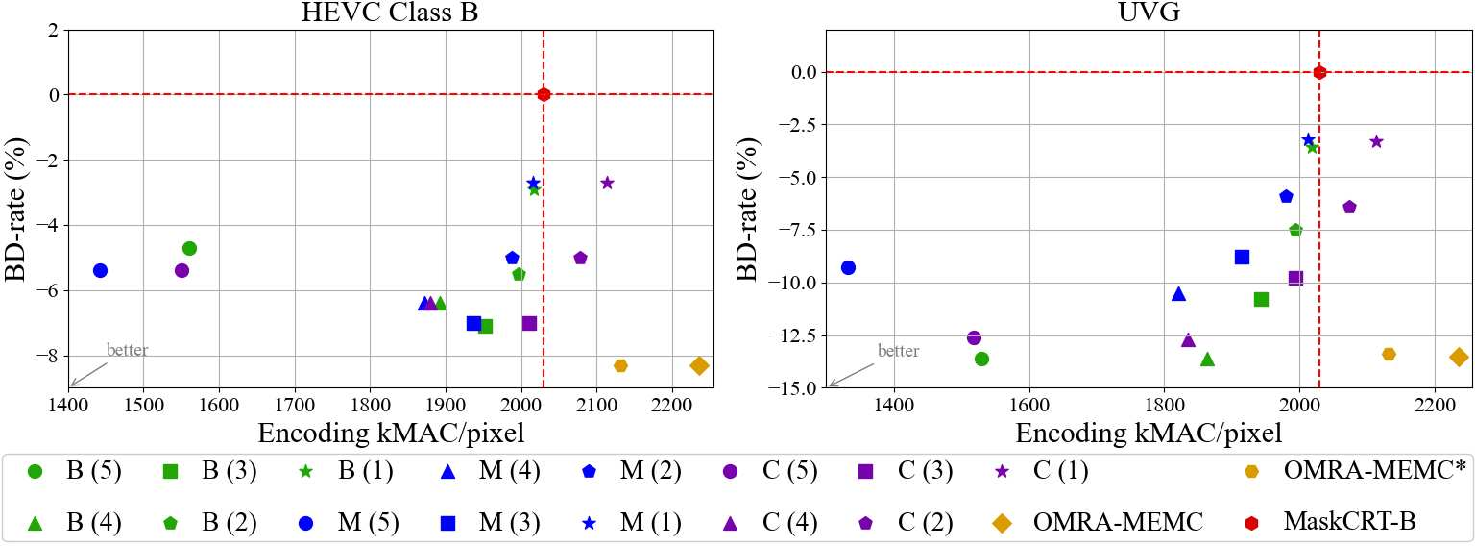}
        \caption{\centering}
        \label{fig:per-layer}
    \end{subfigure}
    \hfill
    \begin{subfigure}{0.33\linewidth}
        \centering
        \includegraphics[width=\linewidth]{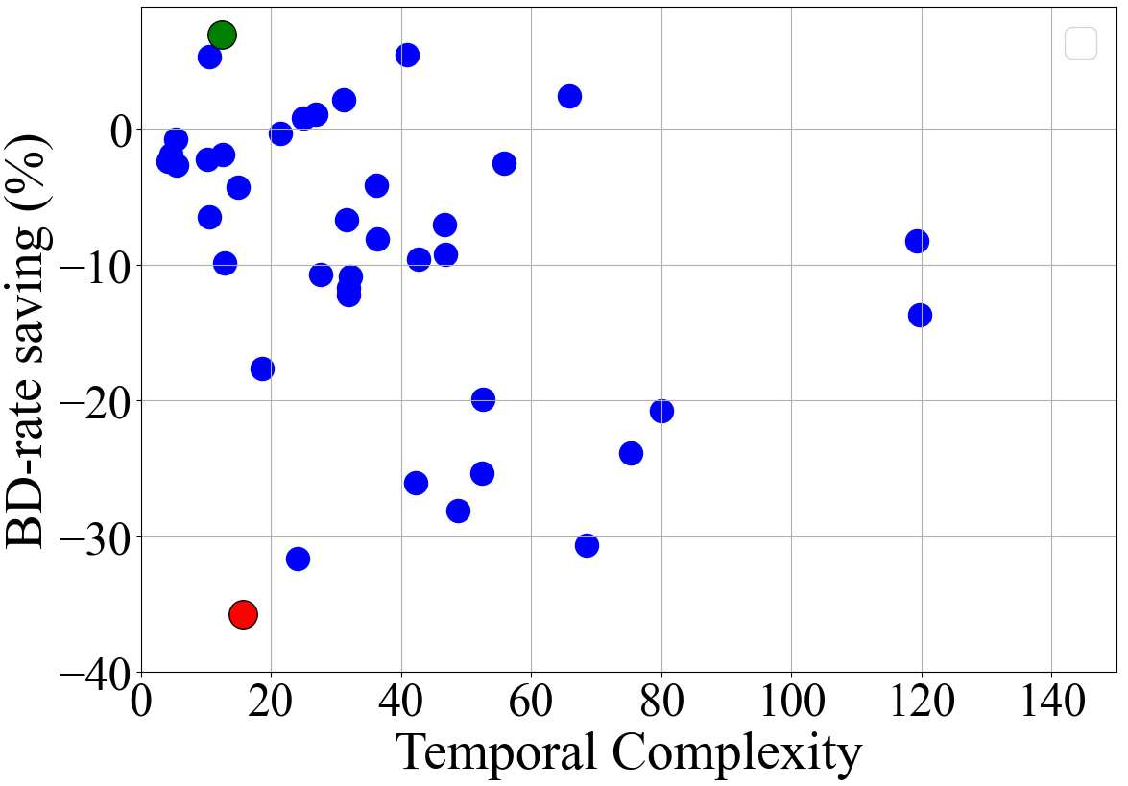}
        \caption{\centering}
        \label{fig:complexity_3dataset}
    \end{subfigure}
    \vspace{-0.8em}
    \caption{(a) The rate-distortion-complexity trade-offs when Bi-Class (B), Mu-Class (M), and Co-Class (C) Fast-OMRA are applied to various temporal layers. The B~(n), M~(n), and C~(n) configurations apply Bi-Class, Mu-Class, and Co-Class Fast-OMRA, respectively, to all video frames in temporal layers up to \textit{n}, and (b) temporal complexity of video sequences versus BD-rate savings achieved by our Co-Class Fast-OMRA. The red and green points indicate the best and worst sequences in terms of BD-rates, respectively.}
    \label{fig:RD_complexity_layer}
    \end{center}
    \vspace{-1.2em}
\end{figure*}

% \input{chapters/table/RD_per_seq}
% \usepackage{multirow}
\begin{table*}[t]
\caption{The per-sequence and/or average BD-rates on the UVG, HEVC Class B, and MCL-JCV datasets, with MaskCRT-B serving as the anchor.}
\centering
\scriptsize
\begin{tabular}{cccccccc}
\hline
\multirow{3}{*}{Dataset} & \multirow{3}{*}{Sequence} & \multirow{3}{*}{OMRA} & \multirow{3}{*}{\begin{tabular}[c]{@{}c@{}}OMRA-\\ MEMC\end{tabular}} & \multirow{3}{*}{\begin{tabular}[c]{@{}c@{}}OMRA-\\ MEMC*\end{tabular}} & \multirow{3}{*}{\begin{tabular}[c]{@{}c@{}}Bi-Class\\ Fast-OMRA\end{tabular}} & \multirow{3}{*}{\begin{tabular}[c]{@{}c@{}}Mu-Class\\ Fast-OMRA\end{tabular}} & \multirow{3}{*}{\begin{tabular}[c]{@{}c@{}}Co-Class\\ Fast-OMRA\end{tabular}} \\
                         &                           &                       &                                                                       &                                                                        &                                                                               &                                                                               &                                                                               \\
                         &                           &                       &                                                                       &                                                                        &                                                                               &                                                                               &                                                                               \\ \hline
\multirow{6}{*}{HEVC-B}  & BasketballDrive           & -22.7                 & -20.4                                                                 & -20.3                                                                  & -17.0                                                                         & -16.2                                                                         & -19.9                                                                         \\
                         & BQTerrace                 & -8.1                  & -7.5                                                                  & -7.4                                                                   & -5.5                                                                          & -8.3                                                                          & -8.1                                                                          \\
                         & Cactus                    & -2.5                  & -1.7                                                                  & -1.7                                                                   & 0.6                                                                           & 0.8                                                                           & 0.8                                                                           \\
                         & Kimono1                   & -12.7                 & -11.2                                                                 & -11.2                                                                  & -12.0                                                                         & -9.3                                                                          & -12.2                                                                         \\
                         & ParkScene                 & -2.7                  & -0.9                                                                  & -0.9                                                                   & 2.1                                                                           & 0.8                                                                           & 2.1                                                                           \\
                         & \textit{Average}          & \textit{-9.7}         & \textit{-8.3}                                                         & \textit{-8.3}                                                          & \textit{-6.3}                                                                 & \textit{-6.4}                                                                 & \textit{-7.5}                                                                 \\ \hline
\multirow{8}{*}{UVG}     & Beauty                    & -2.9                  & -1.7                                                                  & -1.6                                                                   & -1.3                                                                          & -0.6                                                                          & -1.9                                                                          \\
                         & Bosphorus                 & -9.4                  & -6.7                                                                  & -6.7                                                                   & -6.5                                                                          & -5.1                                                                          & -6.5                                                                          \\
                         & HoneyBee                  & -1.4                  & 0.0                                                                   & 0.0                                                                    & -0.8                                                                          & 7.2                                                                           & -0.8                                                                          \\
                         & Jockey                    & -53.0                 & -52.8                                                                 & -52.8                                                                  & -52.8                                                                         & -44.7                                                                         & -47.0                                                                         \\
                         & ReadySteadyGo             & -30.4                 & -29.2                                                                 & -29.2                                                                  & -27.0                                                                         & -25.1                                                                         & -25.4                                                                         \\
                         & ShakeNDry                 & -0.5                  & 0.0                                                                   & 0.0                                                                    & -0.3                                                                          & 1.3                                                                           & -0.3                                                                          \\
                         & YachtRide                 & -7.5                  & -3.8                                                                  & -3.8                                                                   & -6.7                                                                          & -6.6                                                                          & -6.7                                                                          \\
                         & \textit{Average}          & \textit{-15.0}        & \textit{-13.5}                                                        & \textit{-13.4}                                                         & \textit{-13.6}                                                                & \textit{-10.5}                                                                & \textit{-12.7}                                                                \\ \hline
MCL-JCV                  & \textit{Average}          & \textit{-16.0}        & \textit{-11.7}                                                        & \textit{-9.7}                                                          & \textit{-10.5}                                                                & \textit{-9.3}                                                                 & \textit{-10.5}                                                                \\ \hline
\end{tabular}
% \vspace{-1.0em}
\label{tab:RD_per_seq}
% \vspace{-1.5em}
\end{table*}
% \input{chapters/table/trade-offs}

\begin{table*}[h!]
\caption{Trade-offs between BD-rate and encoding complexity of Bi-Class, Mu-Class and Co-Class.}
\centering
\scriptsize
\begin{tabular}{c|cc|cc|cc|cc}
\hline
\multirow{2}{*}{Variant} & \multicolumn{2}{c|}{HEVC-B} & \multicolumn{2}{c|}{UVG} & \multicolumn{2}{c|}{MCL-JCV} & \multicolumn{2}{c}{Average} \\ \cline{2-9} 
                         & BD-rate     & kMAC/pixel    & BD-rate   & kMAC/pixel   & BD-rate     & kMAC/pixel     & BD-rate     & kMAC/pixel    \\ \hline
Bi-Class                 & -6.3        & 1893          & -13.6     & 1863         & -10.5       & 1885           & -10.1       & 1880          \\
Mu-Class                 & -6.4        & 1872          & -10.5     & 1820         & -9.3        & 1822           & -8.7        & 1838          \\
Co-Class                 & -7.5        & 1879          & -12.7     & 1835         & -10.5       & 1833           & -10.2       & 1849          \\ \hline
\end{tabular}
\label{tab:trade_offs}
\vspace{-1.0em}
\end{table*}

\section{Experimental Results}
\label{sec:results}

%\subsection{Test Condition}
%\label{ssec:setting}
The proposed Fast-OMRA variants are integrated into MaskCRT-B and evaluated on HEVC Class B~\cite{hevcctc}, UVG~\cite{uvg}, and MCL-JCV~\cite{mcl} datasets. Following OMRA~\cite{OMRA}, each sequence is encoded with 97 frames, an intra-period of 32, and a GOP size of 32. Encoding complexity is measured in kMAC/pixel and coding efficiency via BD-rate~\cite{bdrate}, with MaskCRT-B as anchor. While we use MaskCRT-B as an example, the proposed Fast-OMRA method is generalizable and can be applied to other learned video codecs and/or other motion estimation networks, as presented in~\cite{OMRA}. We compare our method with OMRA-MEMC and OMRA-MEMC*~\cite{FastOMRA}, which use the same search strategy as~\cite{Murat_ICIP,Murat_OJSP}.

\label{ssec:ablation}
\begin{figure}[t]
    \begin{center}
    \vspace{-1.0em}
    \begin{subfigure}{0.42\linewidth}
        \centering
        \includegraphics[width=\linewidth]{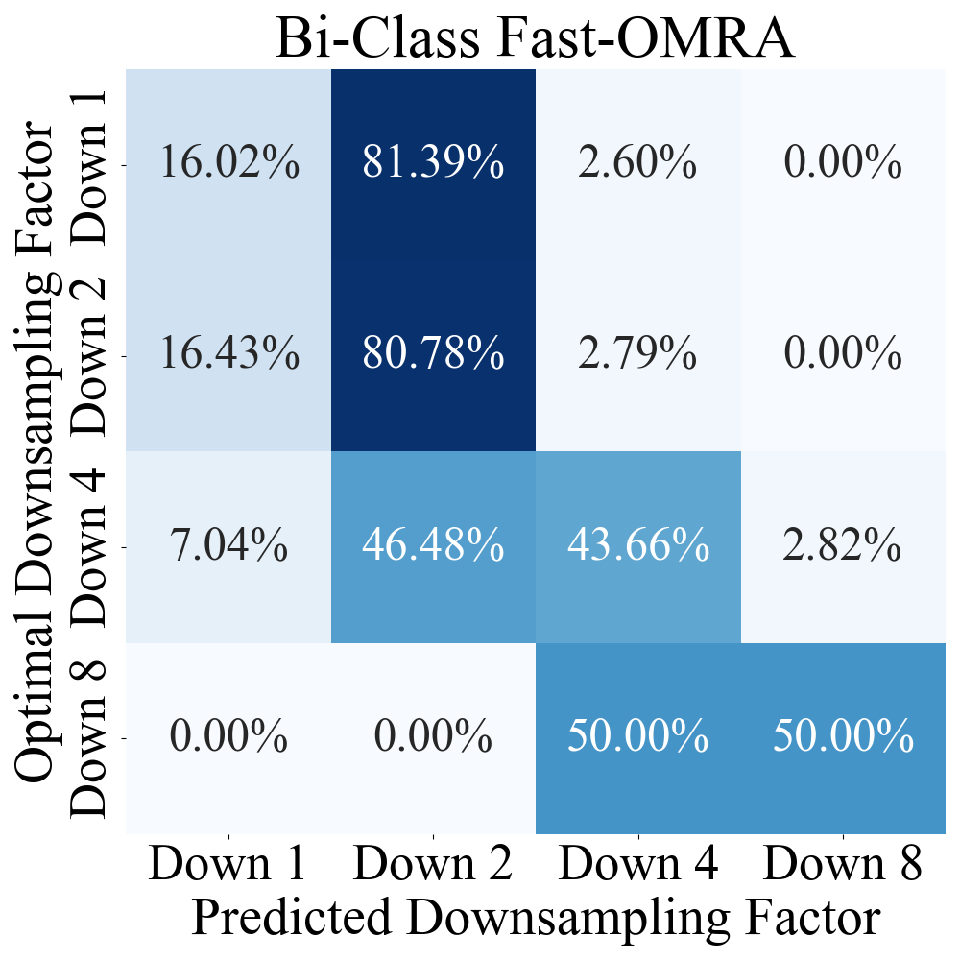}
        \label{fig:biclass}
        \vspace{-1.0em}
    \end{subfigure}
        \hspace{0.5cm}
    \begin{subfigure}{0.42\linewidth}
        \centering
        \centering
        \includegraphics[width=\linewidth]{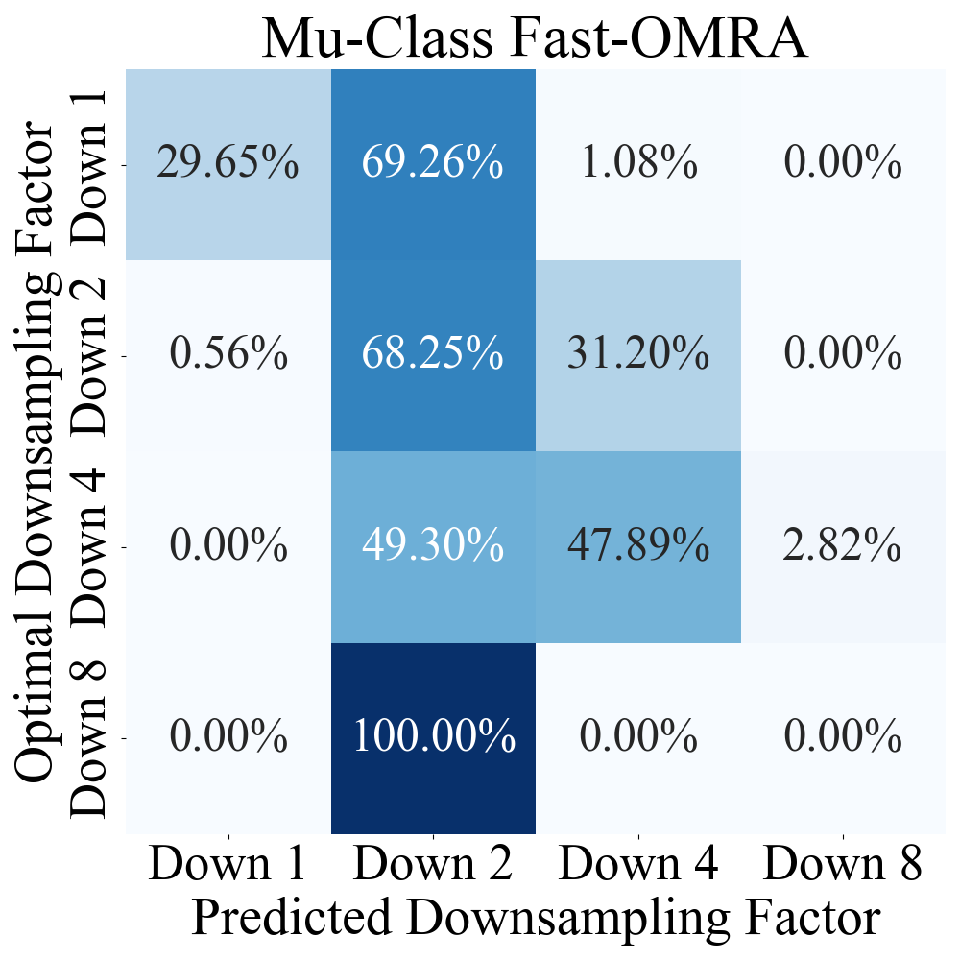}
        \label{fig:muclass}
        \vspace{-1.0em}
    \end{subfigure}
    \begin{subfigure}{0.42\linewidth}
        \centering
        \centering
        \includegraphics[width=\linewidth]{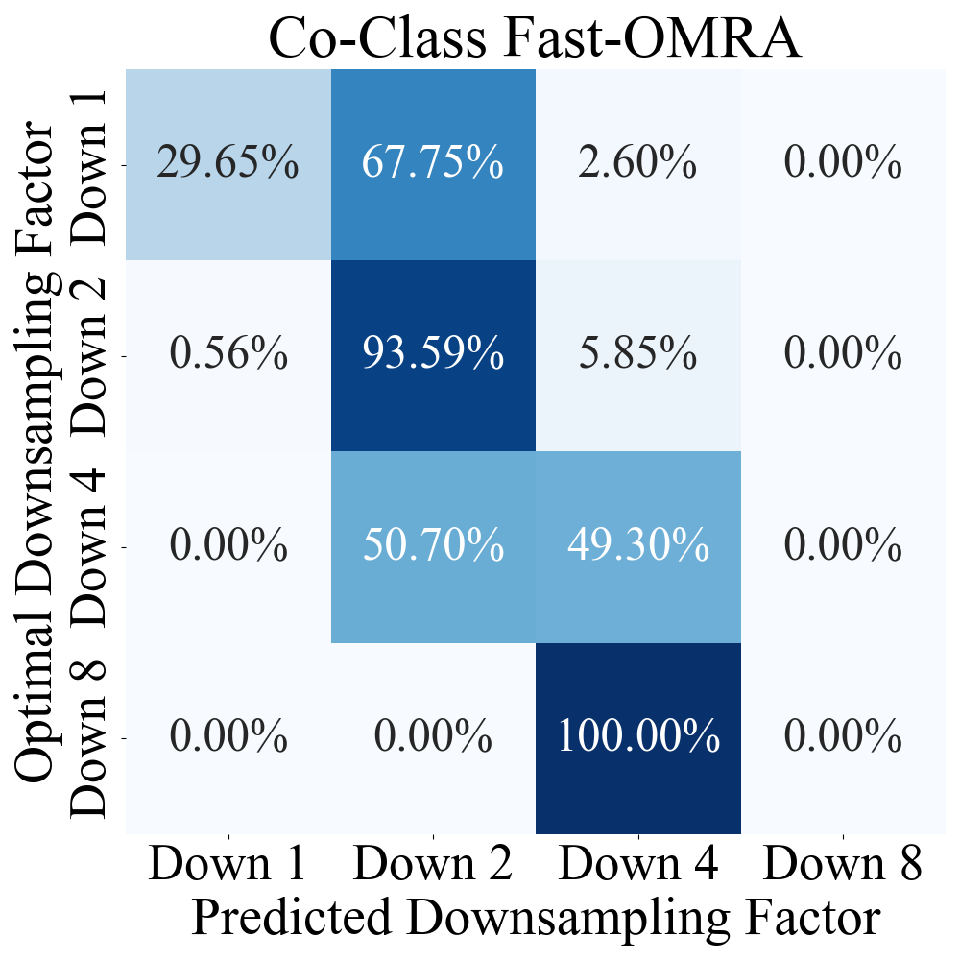}
        \label{fig:coclass}
    \end{subfigure}
    
    \vspace{-1.0em}
    \caption{Confusion matrices of Bi-Class, Mu-Class and Co-Class Fast-OMRA.}
    \label{fig:confusion_matrix_mix}
    \end{center}
    \vspace{-1.9em}
\end{figure}

To evaluate the trade-offs between coding performance and complexity, Table~\ref{tab:RD_per_seq} presents the BD-rate savings for the competing methods, and Table~\ref {tab:trade_offs} compares the trade-offs between BD-rates and encoding kMAC/pixel values across three test datasets. Fig.~\ref{fig:RD_seq} visualizes the RD-curves for BasketballDrive, Jockey, and videoSRC01. The results suggest that our Fast-OMRA schemes (e.g., Bi-Class, Mu-Class, and Co-Class) achieve coding performance comparable to OMRA-MEMC and OMRA-MEMC*~\cite{FastOMRA} while significantly reducing the encoding kMAC/pixel. In some sequences (e.g., BQTerrace, Kimono1, and ShakeNDry), our Fast-OMRA schemes show coding performance similar to OMRA, which achieves the highest coding efficiency but at the substantial computational cost of 5863 kMAC/pixel. By integrating the selective search from Bi-Class, Co-Class outperforms Mu-Class with only minimal computational overhead.

From Table~\ref{tab:RD_per_seq} and Table~\ref{tab:trade_offs}, Co-Class outperforms Mu-Class and achieves compression performance comparable to or better than Bi-Class in some cases, while requiring lower computational cost than Bi-Class with complexity similar to Mu-Class. Fig.~\ref{fig:RD_complexity_layer}(a) evaluates the impact of Fast-OMRA on rate-distortion-complexity trade-offs across temporal layers within a hierarchical B-frame prediction structure. As shown, applying these Fast-OMRA variants to all five temporal layers achieves the lowest encoding kMAC/pixel. However, this approach does not consistently deliver optimal coding performance across datasets. For instance, on the HEVC-B, its performance is inferior to the 4- and 3-layer cases, while limiting to the lowest three layers fails to achieve good efficiency on UVG. The analysis indicates that applying Fast-OMRA to frames in the four lowest temporal layers strikes a favorable balance between encoding complexity and coding performance.

Fig.~\ref{fig:RD_complexity_layer}(b) characterizes the relationship between temporal complexity (computed using the Video Complexity Analyzer~\cite{VCA}) and the BD-rate savings achieved by Fast-OMRA Co-Class across all sequences in the three test datasets. Generally, sequences with higher temporal complexity tend to yield greater BD-rate savings, corroborating Fast-OMRA's ability to mitigate the domain-shift issue.

Fig.~\ref{fig:confusion_matrix_mix} presents the confusion matrices of the proposed Fast-OMRA schemes. The predicted downsampling factors for Bi-Class and Co-Class are obtained after the search process, while Mu-Class predictions are directly generated by the neural network. For fair comparison, identical inputs (i.e., coding frame and reference frames) are used for each B-frame across all schemes, with ground truth determined via exhaustive search for actual encoding.
The results highlight that integrating a search mechanism in Co-Class Fast-OMRA significantly enhances prediction accuracy, particularly for frames requiring downsampling factors of $S = 2$ or $S = 4$.
\vspace{-1.0em}

\section{Conclusion}
This work introduces Fast-OMRA, a lightweight, low-complexity approach designed to mitigate the domain shift issue in neural B-frame coding that adopts hierarchical B-frame prediction. We explore three variants: Bi-Class, Mu-Class, and Co-Class. Bi-Class rapidly determines whether to downsample the coding and reference frames, incorporating additional search steps to determine the downsampling factor. Mu-Class directly outputs the final downsampling factor. Co-Class leverages the strengths of both Bi-Class and Mu-Class, achieving a balanced approach. Experimental results demonstrate that all variants achieve a favorable rate-distortion-complexity trade-off as compared to OMRA. 

% \bibliographystyle{IEEEtran}
% \bibliography{IEEEabrv,paper.bib}

% \onecolumn
% \input{letter_to_editor}

\end{document}